\documentstyle[amsfonts,preprint,aps]{revtex}
\tightenlines
\begin{document}
\title{Poles and zeros of the scattering matrix associated to defect modes}
\author{D. Felbacq}
\address{LASMEA UMR CNRS 6602 \\
Complexe des C\'{e}zeaux\\
63177 Aubi\`{e}re Cedex\\
France}
\maketitle

\begin{abstract}
We analyze electromagnetic waves propagation in one-dimensional periodic
media with single or periodic defects. The study is made both from the point
of view of the modes and of the diffraction problem. We provide an explicit
dispersion equation for the numerical calculation of the modes, and we
establish a connection between modes and poles and zeros of the scattering
matrix.
\end{abstract}

Periodic media with defects have been intensively addressed in the field of
Quantum Mechanics (see \cite{cargese,carmona,figotin} and references
therein) and also, since the development of photonic crystals, in
Electromagnetics \cite{joan,fig,cento,maystre}. On the general subject of
photonic crystals, the interested reader can find an impressive bibliography
on the Internet\cite{biblio}. From the theoretical point of view, the
quasi-totality of the studies are involved in the characterization of the
spectrum of the infinite medium. Nevertheless, for the working physicists,
the main problem is that of the finite structure. Indeed, one can only
imagine experiments, such as the diffraction of a plane wave, by finite
devices. The main question is then to relate the modes with the behavior of
the diffracted field. The simplest connection is that of the determination
of the conduction bands: whenever the frequency belongs to the spectrum of
the infinite structure, the finite one allows the guidance of waves while in
the gap the electromagnetic field decreases exponentially. In the present
paper, we study, as a model problem, the simple case of a one dimensional
periodic structure with defects, which may model for instance a quantum
cavity. Such a device has also been intensively addressed \cite
{lekner,fgz1,fgz2,fig2,leung,liu,fang,bass}. Nevertheless, these studies
involve the infinite structure for which a spectral analysis is given,
whereas in the present communication we aim at establishing a link between
the diffractive properties of a finite structure with a defect, and the
known spectral properties of an infinite structure with a defect.

Throughout this paper, an orthonormal triaxial cartesian coordinate system $%
(0,x,y,z)$ is used. We consider a periodic structure, described by a bounded
real one-periodic function $\varepsilon (x)$, ($\varepsilon (x)=\varepsilon
(x+1)$), representing the relative permittivity with respect to $x$, whereas
the permeability is assumed to be $\mu _{0}$, i.e. that of vacuum. The
structure is assumed to be invariant in the $y$ and $z$ directions and the
harmonic fields (time dependence of $\exp \left( -i\omega t\right) $) are
invariant along $z$. That way, the field is described by a function $%
u_{n}(x)\exp (i\alpha y)$, where $\alpha \in (-\pi ,+\pi ]$ is the Bloch
frequency (in case of a scattering problem, the frequency $\alpha $ is equal
to $k_{0}\sin \theta $, where $\theta $ is the angle of incidence of an
incident plane wave). When the electric field ${\bf E}$ is parallel to the $z
$-axis ($E||$ case), $u_{N}(x)\exp (i\alpha y)$ represents the $z$-component
of ${\bf E}$ and when the magnetic field ${\bf H}$ is parallel to the $z$%
-axis ($H||$ case), it represents the $z$-component of ${\bf H}$. 

Denoting: $\beta ^{2}\left( x\right) =k_{0}^{2}\varepsilon (x)-\alpha ^{2}$, 
$\beta _{0}^{2}=k_{0}^{2}-\alpha ^{2}$, and setting $U=\left(
u,q^{-1}\partial _{x}u\right) $, the propagation equation takes the form:

\begin{equation}
\partial _{x}U=\left( 
\begin{array}{cc}
0 & q \\ 
-q^{-1}\beta ^{2} & 0
\end{array}
\right) U  \label{floquet}
\end{equation}
with: $q\left( x\right) \equiv 1$ for $E||$ polarization, $q\left( x\right)
\equiv \varepsilon \left( x\right) $ for $H||$ polarization. The monodromy
matrix of the equation, or Floquet operator, is the $2\times 2$ matrix ${\bf %
T}_{k,\alpha }$ such that: $U(x+d)={\bf T}_{k,\alpha }U\left( x\right) $.
When considering only $n$ periods of the medium embedded in vacuum and
illuminated by a plane wave (the device extends over $\left[ 0,n\right] $),
the following boundary conditions hold:
\begin{equation}
\left. 
\begin{array}{l}
i\beta _{0}u|_{x=0}+\partial _{x}u|_{x=0}=2i\beta _{0} \\ 
i\beta _{0}u|_{x=n}-\partial _{x}u|_{x=n}=0
\end{array}
\right.   \label{rad}
\end{equation}
from which the reflection and transmission coefficients can be derived: 
\[
\left. 
\begin{array}{l}
r_{n}=u|_{x=0}-1 \\ 
t_{n}=u|_{x=n}
\end{array}
\right. 
\]
In the infinite medium, the conduction bands are characterized by the
condition $\left| tr\left( {\bf T}\right) \right| \leq 2$, we thus define:
\[
\begin{array}{c}
{\bf G}=\left\{ \left( k,\alpha \right) /\left| tr\left( {\bf T}_{k,\alpha
}\right) \right| >2\right\}  \\ 
{\bf B}=\left\{ \left( k,\alpha \right) /\left| tr\left( {\bf T}_{k,\alpha
}\right) \right| <2\right\} 
\end{array}
\]

For $\left( k,\alpha \right) \in {\bf G}\cup {\bf B}$, we denote $\left( 
{\bf v},{\bf w}\right) $ a basis of eigenvectors of ${\bf T}_{k,\alpha }$,
such that $\det \left( {\bf v},{\bf w}\right) =1$, associated to eigenvalues 
$\mu $ and $%
{\displaystyle{1 \over \mu }}%
$( as a convention $\left| \mu \right| <1$ for $\left( k,\alpha \right) \in 
{\bf G}$). We write in the canonical basis of ${\Bbb C}^{2}$: ${\bf v=}%
\left( v_{1},v_{2}\right) ,{\bf w=}\left( w_{1},w_{2}\right) $.

Let us now introduce a defect in the crystal in the following way: we
replace one period of the crystal by a layer of width $h$ and permittivity $%
\varepsilon _{d}\left( x\right) $ ($\varepsilon _{d}$ is simply assumed to
be real and bounded). In view of the spectral problem, the structure is
therefore made of two periodic half-spaces connected by an inhomogeneous
layer extending over $\left[ 0,h\right] $. We denote by ${\bf T}_{0}$ the
monodromy matrix of the defect, omitting here to explicitly write the
dependence in $\left( k,\alpha \right) $. We have the following results:

{\bf Proposition 1(}\cite{carmona,figotin,fig2}{\bf )}: {\it The defect does
not modify the conduction bands}.

{\bf Proposition 2}: \label{local}{\it The defect modes of the structure
correspond to couples }$\left( k,\alpha \right) ${\it \ such that }$\det
\left( {\bf T}_{0}{\bf w},{\bf v}\right) =0$

{\it Proof}: As we introduced a defect, we may now accept an increasing
solution in the left semi-crystal and a decreasing solution in the right
semi-crystal. The problem is to match these two solutions. This is only
possible if ${\bf T}_{0}$ switches increasing solutions to decreasing ones,
that is if ${\bf T}_{0}{\bf w}\in Vect({\bf v})$ QED.

Suppose that the structure is finite and that the defect is switched between 
$n$ periods. The monodromy matrix is: ${\bf T}_{k,\alpha }^{n}{\bf T}_{0}%
{\bf T}_{k,\alpha }^{n}$. In basis $\left( {\bf v},{\bf w}\right) $, ${\bf T}%
_{0}$ is written:
\[
{\bf T}_{0}=\left( 
\begin{array}{cc}
a_{0} & c_{0} \\ 
b_{0} & d_{0}
\end{array}
\right) \text{ .}
\]
Denoting: 
\begin{equation}
\chi =\left( \chi _{ij}\right) =\left( 
\begin{array}{cc}
w_{2}-i\beta _{0}w_{1} & w_{2}+i\beta _{0}w_{1} \\ 
-v_{2}+i\beta _{0}v_{1} & -v_{2}-i\beta _{0}v_{1}
\end{array}
\right) \text{,}  \label{chi}
\end{equation}
an expression of the coefficients $\left( r_{n},t_{n}\right) $ can be easily
obtained: 
\begin{equation}
r_{n}\left( k,\alpha \right) =\frac{p(\mu ^{2n})}{q(\mu ^{2n})},t_{n}\left(
k,\alpha \right) =\frac{-2i\beta _{0}\;\mu ^{2n}}{q(\mu ^{2n})}
\label{reflection}
\end{equation}
where:
\begin{eqnarray*}
p\left( X\right)  &=&\chi _{21}\chi _{11}a_{0}X^{2}+\left( \chi
_{21}^{2}c_{0}-\chi _{11}^{2}b_{0}\right) X-\chi _{21}\chi _{11}d_{0} \\
q\left( X\right)  &=&-\chi _{21}\chi _{12}a_{0}X^{2}+\left( -\chi _{21}\chi
_{22}c_{0}+\chi _{11}\chi _{12}b_{0}\right) X+\chi _{11}\chi _{22}d_{0}\text{
.}
\end{eqnarray*}

\bigskip 

Assume now that there exists some defect mode for a couple $\left(
k_{0},\alpha _{0}\right) $ so that $\left| \alpha _{0}\right| <k_{0}$,
allowing to define an angle of incidence $\theta _{0}$ by $\alpha
_{0}=k_{0}\sin \theta _{0}$. 

It is easily seen that the equation of Proposition 2 simply writes: $%
d_{0}\left( k_{0},\alpha _{0}\right) =0$ , whence we get from (\ref
{reflection}): 
\begin{eqnarray*}
r_{n}\left( k_{0},\theta _{0}\right)  &=&\frac{-\chi _{11}^{2}b_{0}+\chi
_{21}^{2}c_{0}+\chi _{21}\chi _{11}\mu ^{2n}a_{0}}{\chi _{11}\chi
_{12}b_{0}-\chi _{21}\chi _{22}c_{0}-\chi _{21}\chi _{12}\mu ^{2n}a_{0}} \\
t_{n}\left( k_{0},\theta _{0}\right)  &=&\frac{-2i\beta _{0}}{-\chi
_{21}\chi _{22}c_{0}+\chi _{11}\chi _{12}b_{0}-\chi _{21}\chi _{12}\mu
^{2n}a_{0}}
\end{eqnarray*}
As $n$ tends to infinity, we have the following limits: 
\begin{equation}
\begin{array}{ll}
r_{n}\left( k_{0},\theta _{0}\right) \longrightarrow 
{\displaystyle{\chi _{21}^{2}c_{0}-\chi _{11}^{2}b_{0} \over \chi _{21}\chi _{11}b_{0}-\chi _{22}\chi _{21}c_{0}}}%
& t_{n}\left( k_{0},\theta _{0}\right) \longrightarrow 
{\displaystyle{-2i\beta _{0} \over \chi _{21}\chi _{11}b_{0}-\chi _{22}\chi _{21}c_{0}}}%
\\ 
r_{n}\left( k,\theta _{0}\right) \longrightarrow -%
{\displaystyle{\chi _{21} \over \chi _{22}}}%
\text{, for }k\neq k_{0} & t_{n}\left( k,\theta _{0}\right) \longrightarrow 0%
\text{, for }k\neq k_{0}
\end{array}
\label{limitr}
\end{equation}
{\bf Remark 1}: Obviously from (\ref{chi}): $\left| -%
{\displaystyle{\chi _{21} \over \chi _{22}}}%
\right| =1$, and so as a conclusion $\left| r_{n}\left( k_{0},\theta
_{0}\right) \right| $ tends pointwise towards a value that is strictly less
than $1$, whereas for any point different from $k_{0}$, in a small enough
neighborhood of $k_{0}$, $\left| r_{n}\left( k,\theta _{0}\right) \right| $
tends to $1$ as $n\longrightarrow \infty $. This result means that the
reflected energy admits a sharp minimum near $k_{0}$. It is important to
note that the minimum of $\left| r_{n}\left( k,\theta _{0}\right) \right| $
is not {\it a priori} reached for value $k_{0}$ of the wavenumber, but the
sharpness is all the more important as $n$ is important. Clearly, for a
given incidence $\theta _{0}$, it is possible to transmit waves of
wavenumber belonging to {\it a small interval near}{\bf \ }$k_{0}$ and
tending to $\left\{ k_{0}\right\} $ as $n$ tends to infinity{\it \ }

It is known that for a fixed incidence $\theta $, the scattering matrix
admits a meromorphic extension to the complex half-space $\left\{ 
\mathop{\rm Im}%
(k)<0\right\} $\cite{reed}. From (\ref{reflection}), poles and zeros of $%
r_{n}$ are solutions respectively of equations: 
\begin{equation}
\begin{array}{r}
\mu ^{4n}a_{0}+\mu ^{2n}\left( 
{\displaystyle{\chi _{21} \over \chi _{11}}}%
c_{0}-%
{\displaystyle{\chi _{11} \over \chi _{21}}}%
b_{0}\right) =d_{0} \\ 
\mu ^{4n}%
{\displaystyle{\chi _{21}\chi _{12} \over \chi _{11}\chi _{22}}}%
a_{0}+\mu ^{2n}\left( 
{\displaystyle{\chi _{21} \over \chi _{11}}}%
c_{0}-%
{\displaystyle{\chi _{12} \over \chi _{22}}}%
b_{0}\right) =d_{0}
\end{array}
\label{pozo}
\end{equation}
and the limit of both equations as $n$ tends to infinity is just $%
d_{0}\left( k,k\sin \theta \right) =0$, which is the equation defining the
defect modes. As both equations (\ref{pozo}) are analytic perturbations of $%
d_{0}\left( k,k\sin \theta \right) =0$ we can conclude:

{\bf Proposition 3:} {\it One defect mode }$\left( k_{0},k_{0}\sin \theta
_{0}\right) ${\it \ is associated to exactly one zero }$k_{z}^{n}${\it \ of }%
$r_{n}${\it \ and one pole }$k_{p}^{n}${\it \ of }$r_{n}${\it \ and }$t_{n}$%
{\it \ . Moreover }$k_{z}^{n}${\it \ and }$k_{p}^{n}${\it \ tend to }$k_{0}$%
{\it \ as }$n${\it \ tends to infinity.}

{\bf Remark 2}: This suggests that, for sufficiently large $n$, there exist
two positive numbers $\gamma _{n}$ and $\delta _{n}$ such that: 
\[
\begin{array}{l}
k_{z}^{n}=k_{0}-i\gamma _{n}\delta _{n}+O\left( \delta _{n}\right)  \\ 
k_{p}^{n}=k_{0}-i\delta _{n}+O\left( \delta _{n}\right) 
\end{array}
\]
We know that, as $n$ tends to infinity, $\delta _{n}$ tends to $0$ and that $%
r_{n}\left( k_{0},\theta _{0}\right) $ admits a limit given by (\ref{limitr}%
). In the vicinity of $k_{0}$, we can write $r_{n}\left( k,\theta \right)
\simeq -%
{\displaystyle{\chi _{21} \over \chi _{22}}}%
{\displaystyle{k-k_{z}^{n} \over k-k_{p}^{n}}}%
$ whence we obtain: $r_{n}\left( k_{0},\theta _{0}\right) \simeq -%
{\displaystyle{\chi _{21} \over \chi _{22}}}%
{\displaystyle{%
\mathop{\rm Im}\left( k_{z}^{n}\right)  \over %
\mathop{\rm Im}\left( k_{p}^{n}\right) }}%
$, so that $\gamma _{n}=%
{\displaystyle{\chi _{11}^{2}\chi _{21}^{-1}b_{0}-\chi _{21}c_{0} \over \chi _{21}c_{0}-\chi _{21}\chi _{11}\chi _{22}^{-1}b_{0}}}%
$.

{\bf Remark 3}: For real $z$, $z\mapsto -%
{\displaystyle{\chi _{21} \over \chi _{22}}}%
{\displaystyle{z-k_{z}^{n} \over z-k_{p}^{n}}}%
$\ is the equation of a circle of diameter $\sqrt{1+\gamma _{n}^{2}}$, so
that while in the gap the reflection coefficient belongs to the unit circle
of the complex plane, it describes a circle for real $k$\ varying in the
vicinity of $k_{0}$.

In order to be able to apply Bloch-waves method to defect structures, an
approximate way consists in periodizing the defect, it is the so-called
supercell approximation. Let us apply this method to our simple example. We
consider thus an infinite periodic structure whose period is made of one
defect switched between $n$ periods of the previous medium. We call it a
super-structure. We know that the global monodromy matrix of the
super-structure is: $\widetilde{{\bf T}}_{k,\alpha }={\bf T}_{k,\alpha }^{n}%
{\bf T}_{0}{\bf T}_{k,\alpha }^{n}$, and that the conduction bands are
characterized by $\left| tr\left( \widetilde{{\bf T}}\right) \right| <2.$
Now what happens at $k=k_{0}$? Obviously:$tr\left( \widetilde{{\bf T}}%
_{k,\alpha }\right) =\mu ^{2n}a_{0}+\mu ^{-2n}d_{0}$ so that for $\left(
k,\alpha \right) =\left( k_{0},\alpha _{0}\right) $ : $tr\left( \widetilde{%
{\bf T}}_{k_{0},\alpha _{0}}\right) =\mu ^{2n}a_{0}$ and thus tends to $0$
as $n$ tends to infinity. This means that, at $\alpha =\alpha _{0},$ there
exists an interval $J_{n}$ of wavenumbers $k$ over which $\left| tr\left( 
\widetilde{{\bf T}}_{k,\alpha _{0}}\right) \right| <2$ and therefore $J_{n}$
is a conduction band in the super-structure. Of course the length of this
interval depends upon $n$ and tends to $0$ as $n$ tends to infinity because $%
J_{n}$ $\ $tends to $\left\{ k_{0}\right\} $. This is a crucial point as it
justifies the use of Bloch-waves theory in the framework of the super-cell
approximation, {\it at least in a small neighborhood of }$k_{0}$. Finally,
we can state:

{\bf Proposition 4}: {\it In the super-structure, the existence of a defect }%
$\left( k_{0},\alpha _{0}\right) ${\it \ implies the opening of conduction
bands inside gaps of the unperturbed structure that can support defect
modes. The widths of these ''defect''-conduction bands tend to zero
exponentially with the number }$n${\it \ of subperiods constituting one
period of the super-structure.}

{\bf Remark 4}: For a finite superstructure, and as the medium is periodic,
we may apply a theory \cite{fgz2} allowing to compute the superior envelop
of the modulus of the reflection coefficient of a periodic structure. In the
present problem, this is just the graph of: 
\[
k\longmapsto \sqrt{1-\left( 4-tr(\widetilde{{\bf T}})^{2}\right) \left( 
\widetilde{{\bf t}}_{12}\beta _{0}-%
{\displaystyle{\widetilde{{\bf t}}_{21} \over \beta _{0}}}%
\right) ^{-2}}
\]
where $\widetilde{{\bf T}}=\left( \widetilde{{\bf t}}_{ij}\right) .$ 

We have analyzed wave propagation in one-dimensional photonic crystals with
one defect. We have shown a connection between the scattering properties and
the modes: defect modes of the infinite structure give rise to a pole and a
zero explaining the behavior of the refection coefficient. These results
should help in the understanding of more complicated structures such as
bidimensional photonic crystals, for which our present work could be used as
an approximate theory \cite{cento,maystre}$.$

\end{document}